\documentstyle[12pt]{article}

\begin{document}

\begin{flushright}
math-ph/0105044
\end{flushright}

\vspace{5mm}

\begin{center}
{\Large \sf Solitons of the Self-dual Chern-Simons Theory on a
Cylinder}\\[10mm]
{\sf Seongtag Kim}\footnote{Electronic mail: stkim$@$skku.ac.kr} \\
{\it Department of Mathematics,
Sungkyunkwan University,\\
Suwon 440-746, Korea}
\end{center}

\vspace{5mm}

\begin{abstract}
We study the self-dual Chern-Simons Higgs theory on an
asymptotically flat  cylinder. A topological multivortex
solution  is constructed and the fast decaying property of
solutions is proved.
\end{abstract}

\newpage
\def\ep{\epsilon}
\def\bea{\begin{eqnarray}}
\def\eea{\end{eqnarray}}
\def \bu{\overline{u}}
\def \BB{(2.22)}
\def\om{\Omega}
\def\bg{\overline{G}}

\section{Introduction}
\vspace*{-0.5pt}
 \noindent
 In this paper, we prove the existence
and the fast decaying property of topological Chern-Simons
vortices on $(2+1)$ space $(R \times M,~dt^2-g_{ij} dx^i dx^j)$
whose spatial manifold $M$ is a cylinder. The spatial metric
$g_{ij}$ on $M$ is asymptotically flat, so that  each end of the
cylinder is close to $\left(R^2-B_E(r_0), \delta_{ij}\right)$,
the outside of a large ball of radius $r_0$ in Euclidean space
$R^2$. The metric on cylinder $M$ is given in a general form. For
an example, it includes the wormhole of Einstein-Rosen bridge,
which connects two asymptotically flat universes. For the Higgs
potential term, we take the model developed on $(2+1)$ space $(R
\times R^2, dt^2-dx_1^2-dx_2^2)$ by Hong-Kim-Pac
 and Jackiew-Weinberg
to study vortex solutions of the Abelian Higgs model  carrying
both electric and magnetic charges \cite{HKP,JW}.

In the context of recent physics topics including string theory
and brane world scenario, asymptotic cylindrical geometry formed
by topological defects attracts attention. When the smooth brane
world of Randall-Sundrum type~\cite{RS} is considered with
extra-dimensions, the bulk topological defects form their
asymptotic space which is  a cylinder of finite  neck
~\cite{Gre}. In particular, a crucial role of angular momentum
carrying self-dual extended objects is clear in making a stable
(1/4)-supersymmetric tubular D2-brane~\cite{MT}. Since the
solitons of our interest are self-dual and spinning, which live
in a cylindrical background spacetime, they can be viewed as one
candidate of 0-brane counterparts and their worldvolume
realization may become a sort of supertubes or brane worlds.

Over the past decade, many attention have been given to the above
model \cite{HKP, JW} on  spatial manifold $R^2$. For the
Chern-Simons Higgs theory on flat $R^{2}$
 a topological multi-vortex solution and a nontoplogical multi-vortex
 solution exist \cite{Wan}-\cite{CI}.
  Inspired by the work of Hoopt, periodic
solutions on a torus or a sphere were studied
\cite{Hf}-\cite{DJLW3}. Schiff \cite{Sch} constructed a
background metric model on spatial manifold $R^2$ and studied the
radial symmetric case. Generalizations to non-radial background
metric models on the spatial manifold $R^2$ were studied for
nontopological solutions and topological solutions \cite{Choe,
KK}. Recently the topology of the configuration space of this
model on $R^2$ and a cylinder is  analyzed \cite{FG} and an
Abelian gauge field theory on complex line bundle over a compact
surface is developed \cite{ssy}.

Though one of the final goals in this direction is to study the
existence of self-gravitating multi-soliton configurations,
present researches are limited to those with rotational symmetry
due to complications ~\cite{Cle2,cckk}.

In the previous work \cite{Wan}, the distance function is used for
the existence proof  of solitons on $(2+1)$ space $(R \times R^2,
dt^2-dx_1^2-dx_2^2)$. Since the distance function is no longer
smooth on cylinder $M$, we construct an approximate solution
using the Green function on small sets around the centers of
vortices to show the existence.   The Maximum principle is used
to estimate the asymptote of solitons.

\section{ Chern-Simons Equation}  \noindent
In this section we  review the  Bogomolnyi bound of the
Chern-Simons Higgs theory coupled to background gravity
\cite{Sch}.  The static metric $G_{\mu\nu}$ on $(2+1)$ space
$R\times M$ is given by
\begin{equation}
ds^{2}= dt^{2}-g_{ij}(x^{k})dx^{i}dx^{j} ~~~ (i,j,k,... =1,2),
\end{equation}
where $g_{ij}$ is a metric on a two-dimensional cylindrical
spatial manifold $M$. Throughout this paper, we  assume that there
exists a smooth compact subset $K \subset M$ such that $(M-K,
g_{ij})$ has two disjoint connected components $C_1$ and $C_2$,
where $C_1=\left(R^2-B_E(0, R_0), h_1 \delta_{ij}\right)$ and
$C_2=\left(R^2-B_E(0, R_0), h_2 \delta_{ij}\right)$. In the
above, $R^2-B_E(0, R_0)$ denotes the complement of the Euclidean
ball of radius $R_0$ whose center at the origin in $R^2$, and $
h_1$ and $h_2$ are smooth positive functions on $R^2-B_E(R_0)$
satisfying $\alpha^{-1}<h_1, h_2<\alpha$ for some positive
constant $\alpha$.
 The static scalar potential
$V(|\phi|)$ is taken to be
\begin{equation}\label{pot}
V(|\phi|)=\frac{e^{4}}{\kappa^{2}}|\phi|^{2}(|\phi|^{2}-v^{2})^{2}.
\end{equation} \noindent
 The action is given by
\begin{equation}
S=\int d^{3}x\sqrt{G}\left[
\frac{\kappa}{4}\frac{\epsilon^{\mu\nu\rho}}{\sqrt{G}}  F _{\mu
\nu }A_{\rho} + G^{\mu\nu}\overline{D_{\mu}\phi}
D_{\nu}\phi-V(|\phi|) \right],
\end{equation}
where $\phi=e^{i\Theta}|\phi|$ is a complex scalar field,
$A_{\mu}$ a U(1) gauge field,  $F_{\mu\nu}=\partial_{\mu} A_{\nu}-
\partial_{\nu} A_{\mu}$, $D_{\mu}=\partial_{\mu}-ieA_{\mu}$,
and $\sqrt{G}=\sqrt{det(G_{ij})}$
$\left(\sqrt{g}=\sqrt{det(g_{ij})}\right)$. All components are
assumed to be static.  The symmetric energy-momentum tensor is
\begin{equation}
T_{\mu\nu}= (\overline{D_{\mu}\phi}D_{\nu}\phi+
\overline{D_{\nu}\phi}D_{\mu}\phi)-G_{\mu\nu}\left[ G^{\rho\sigma}
\overline{D_{\rho}\phi}D_{\sigma}\phi-V(|\phi|)\right].
\end{equation} \noindent
Since  a Riemann surface admits an isothermal coordinates on a
sufficiently small neighborhood of  each point,
 the metric can be written as $g_{ij}= h(x_1, x_2)~\delta_{ij}=
  \sqrt{g}~\delta_{ij}$ on this
 neighborhood for a positive function $h$. From now on, we use this coordinate system.
 The equation of motion with respect to $A_0$ is given by:
 \bea A_0=-{ {k F_{12}}\over { 2 \sqrt{g} e^2 |\phi|^2}}. \label{a0} \eea
 Assume that $\phi$ is integrable in the following Eq.~(\ref{e45}), \bea E &=& \int_M
T_{00}~dV_g \label{e45}
\\ &=& \int_M d^2 x \left|
\frac{k F_{12}} {2 \sqrt{h} e \phi} \mp \frac{\sqrt{h} e^2}k
\overline{\phi} (|\phi|^2-v^2) \right|^2+ \left|(D_1\pm iD_2) \phi
\right|^2 \mp \int_M d^2 x e F_{12} \nonumber \eea \noindent
where $ dV_g = \sqrt{g}~d^2x$. \noindent
  The first-order Bogomolnyi equations are
\begin{equation}\label{bog1}
F_{12}=\pm\frac{2 e^{3}}{\kappa^{2}}|\phi|^{2}(|\phi|^{2}-v^{2})
\sqrt{g},
\end{equation}
\begin{equation}\label{bog2}
D_{1}\phi\pm i D_{2}\phi=0.
\end{equation}
The second equation (\ref{bog2}) expresses the spatial components
of the gauge field  $A_i$ in terms of the scalar field, i.e.,
$eA_{i}=-\partial_{i}\Theta \mp\ep_{ij} \partial_{j}\ln |\phi|$.
Substituting it into the first Bogomolnyi equation (\ref{bog1})
together with the conformal gauge, we have
\begin{equation}\label{boge}
\partial^{2}\ln
|\phi|=\frac{2 e^{4}}{\kappa^{2}}\sqrt{g}
|\phi|^{2}(|\phi|^{2}-v^{2})
\mp\epsilon^{ij}\partial_{i}\partial_{j}\Theta ,
\end{equation}
where Dirac-delta function like contribution of the scalar phase
$\Theta$ comes from multi-valued function. We define: \bea \tilde
{F_{12}} \equiv { {eF_{12}}  \over {\sqrt{g}} } &=&\pm
{{\partial^2 \ln |\phi|^2}\over {2 \sqrt{g}}} \nonumber  \\&=&
\pm {{2 e^4v^4}\over{k^2}}\left|{\phi\over v}\right|^2 \left(
\left|{\phi\over v}\right|^2 -1 \right ). \label{sum} \eea

To make energy finite, $|\phi|$ can have four possible asymptote,
i.e. $|\phi|\to v $, or zero at the infinity of $C_1$ or $C_2$. A
solution of Eq.~(\ref{boge}) is called a topological solution if
$|\phi|\to v\not=0$ at the infinity of $C_1$ and $C_2$. In this
paper, we consider the topological solution only.

\section{ Existence and Behavior of a Solution}  \noindent
In this section, we show that a topological solution with
arbitrary prescribed vortex centers can be constructed and the
solution  decays fast. The  proof is different from the case of
spatial manifold $R^2$, where $\ln (x_1 ^2+x_2 ^2)$ is used to
construct a solution \cite{Wan}.  The main reason is that the
distance function is not smooth on cylinder $M$. There are at
least two points on $M$ where the distance function is not
differentiable.

From now on, we denote that $M$ be a cylinder, $d(p,q)$ be the
Riemannian distance between two points $p$ and $q$ on $(M,g)$ and
$B(p,r)=\{q\in M~| ~d(p,q)<r~\}$.   Using the isothermal
coordinates system, we let
 $\Delta ={1 \over {\sqrt {g}}}
({{\partial^2}\over{\partial x_1^2}} +{{\partial^2}\over{\partial
x_2^2}})$  ($\Delta_0 ={{\partial^2}\over{\partial  x_1^2}}
+{{\partial^2}\over{\partial x_2^2}}$), $|\nabla u| (  |\nabla
u|_{E})$ and  $\delta $ ($\delta_{E} $) be the Laplacian, the norm
of the gradient and Dirac-delta function with respect to the
metric $g_{ij}$ (Euclidean metric), respectively. We denote
$\parallel w\parallel_p=\left (\int_M |\nabla w|^p
~dV_g\right)^{1\over p}$ and $ H_1^2 $ be the Sobolev space,
which is the completion of $C_c^{\infty}(M)$ with respect to  the
norm $||w||=\big( \int_{M} |\nabla w|^2 +w^2 \ dV_{g})^{1\over
2}$.

 Note that Eq.~(\ref{sum})  is equivalent to Eq.~(\ref{c1}) in the following
Theorem 1 by taking $\left|{\phi\over v}\right|^2=e^w$ and $k=2
e^2 v^2$. We state the main Theorem.

 \vspace*{12pt}\noindent {\bf
Theorem~1.} There exists a topological solution for the following
self-dual Chern-Simons vortex equation on $(M,g)$
\begin{equation}
 \Delta w= e^w (e^w-1) +  4 \pi \sum _{k=1}^{n}\delta_{p_k}(p)
 \label{c1}
\end{equation}
\noindent with the boundary condition \[ \lim_{d(p, p_0) \to
\infty} w=0,
\] \noindent for some
fixed point $p_0 \in M $. Moreover, $w$ satisfies $ -a
e^{-b~d(p,p_0)} \le w(p)<0$  and $|\nabla w|(p)< a
e^{-b~d(p,p_0)}$ at infinity for some positive constants $a$ and
$b$.

\vskip 0.3 true cm \it Remark. \rm  The quantity $\left|{\phi\over
v}\right|^2-1, ~ |D_1 \phi|^2+|D_2 \phi|^2$, and
$\tilde{F_{12}}={{F_{12}}\over{\sqrt{g}}}$ all decay exponentially
fast. This fact easily comes from Theorem 1.

 \vspace*{12pt} \noindent
 {\bf  Proof.}  We divide the proof into two steps.\par
 Step 1. {\it Existence of a solution.} \par \noindent
 We construct a  solution $w$ with $w \to -\infty$ around centers
 of vortices. It is easy to see that any $H_1^2$ solution $w$ of
 Eq.~(\ref{c1}) satisfies
 $w\le 0$ on the outside of centers of vortices, by multiplying $u$ on Eq.~(\ref{c1}) and integrating over
 a large smooth subdomain of $M$.
Let $\{p_1, \cdots p_n  \}$ be the arbitrary prescribed centers of
vortices on $M$, which may not be distinct. Take
  $i_0$ be the injective radius of $(M,g)$,
which means that $B(p,i_0)$  is diffeomorhpic to an open ball in
Euclidean space $(R^2, \delta_{ij})$. Denote $\ep_1= \{1, i_0/2,
d(p_i,p_j)/4 ~|\mbox{ for}~ p_i\neq p_j \} $ and  $
L_0=\max\limits_{k=1,\cdots n} \{4, 4 d(p_1,p_k) \} $.  We
decompose $M=\om_1\cup\om_2\cup \om_3 $, where \bea \om_1 &=&
\{p\in M| \min\limits_{k=1,\cdots n} d(p,
p_k) \le \ep_1\},\nonumber \\ \om_2 &=&M-B(p_1,L_0), \nonumber\\
\om_3 &=& M-(\om_1 \cup \om_2) . \eea \noindent
 By the existence of a Green function on a Riemannian
manifold  \cite{Au}, there exists $u_k $ on $B(p_k, \ep_1)$ for
$k=1,\cdots n$, satisfying the following properties \bea \Delta
u_k(p) =4 \pi m_k \delta_{p_k} (p) ~~\mbox{when} ~~  p \in
B(p_k,\ep_1), \eea \noindent and \bea u_k(p) =\ln \ep_1^{2m_k} ~~
\mbox{when} ~~    p \in
\partial B(p_k,\ep_1), \eea \noindent where $m_k$ is the
multiplicity at $p_k$.
 We take $\overline{u}$ be a smooth function on $M$ with
\[\overline{u}(p)= \left\{
\begin{array}{ll}
 u_k
& \mbox{if }
  x\in B(p_k, \ep_1) ~~  \mbox {for } ~k=1,\cdots n  ,  \nonumber\\
 0
& \mbox{if }
  x \in \om_2,
\end{array}
\right. \] \noindent and  $ ~ \ln \ep_1^{2n}\le \overline{u}(p)
\le 0$
  for   $p \in \om_3 $.
We define $S(p)\equiv e^ {\overline{u}} (p)$.  The metric on
$B(p_k,r)$ approaches  to $\delta_{ij}$ as $r\to 0$ in the normal
coordinates  and $S(p) \to d(p,p_k)^{2m_k}$ as $p \to p_k$
\cite{Au}. Note that $S(p)=1$ on $p\in\om_2$. Take
$w=\overline{u}+u$, then Eq.~(\ref{c1}) becomes
\begin{equation}
 \Delta u = S e^u (S e^u-1) +  h,    \label{c9}
\end{equation}
\noindent where $h=-\Delta \overline{u} +4 \pi \sum
_{k=1}^{n}\delta_{p_k}(p)$ is a smooth function whose support
lies in a compact set $\om_3$.
 A critical
point of functional $E$ defined on
 $ H_1^2$ is a solution
of Eq.~(\ref{c9}), where \bea E(u)=
  \int_{M} |\nabla u|^2 +(
S e^u-1)^2+2 h u \, ~dV_{g} . \label{c10} \eea
 \noindent
 Using the basic inequality  $ (e^t-1)^2 \ge { |t|^2 \over{ (1+
|t|)^2}} $, we estimate the second term of Eq.~(\ref{c10}) \bea
\int_{M} (S e^u-1)^2 dV_{g} &=& \int_{M-B(p_1,L_0)} (e^u -1)^2
~dV_g  +\int_{B(p_1,L_0)} (S e^u-1)^2 ~dV_{g} \nonumber\\
&\ge&  \int_{M-B(p_1,L_0)} (e^u -1)^2 ~dV_g \nonumber\\
&\ge&  \int_{M}  { |u|^2 \over{ (1+ |u|)^2}} ~dV_g -c_1
\label{c12}\eea \noindent  for $u\in H_1^2$, where $c_1$ is the
volume of $B(p_1, L_0)$.

\noindent Since $(M,g)$ is an asymptotically flat cylinder, the
Sobolev Imbedding Theorem holds on $(M, g)$, i.e., there exists a
positive constant $c_2$ such that $\int_{M}f^2 ~dV_g \le
{c_2\over 4} \Big( \int_{M} |\nabla f| ~dV_g \Big)^2  $ for $f
\in H_1^1(M)$ \cite{Au}. Following \cite{Wan, Yi}, we estimate
the last term in (\ref{c10}).
 Set $ f=u^2$, then \bea
 \int_{M} u^4 ~dV_g &\le& c_2\left( \int_{M} | u \nabla u|
dV_{g}\right)^2
\nonumber   \\
 &\le&  c_2\int_{M}  u^2 ~dV_g    \int_{M}  |\nabla
u|^2 ~dV_g , \label{c31}  \eea \noindent and
 \bea
 \left(\int_{M} u^2 ~dV_g \right)^2 &\le &
  \left[ \int_{M}  \left(    {|u| \over { 1+ |u|}} \right) (
1+|u|) |u|   ~dV_g \right]^2 \nonumber   \\
  &\le&  2 \int_{M} \left(  {  |u| \over {
1+|u|} } \right)^2 ~dV_g ~
 \int_{M} u^2 + u^4   ~dV_g  . \label{c34} \eea
\noindent
 From Eqs.~(\ref{c31}) and (\ref{c34}),
 \bea
 \int_{M} u^2 ~dV_g  \le
2 \int_{M} \left(  {  |u| \over { 1+|u|} } \right)^2 ~dV_g \
\left( 1+ c_2 \int_{M} |\nabla u|^2 ~dV_g  \right). \label{c344}
\eea Using $\sqrt{2ab} \le |a|+|b|$,
 \bea \left(\int_{M} u^2 ~dV_g \right)^{1 \over 2} \le
\int_{M} { |u|^2 \over{ (1+ |u|)^2}} ~dV_g + c_2 \int_{M} |\nabla
u|^2 ~dV_g +1. \label{c27} \eea \noindent We can bound  the last
integral in Eq.~(\ref{c10}):
 \bea 2 |\int_M h u ~dV_g| &\le& 2 \parallel h \parallel_{4 \over 3}
\parallel u \parallel_{4} \label{c318}\\
&\le& c_3 \parallel h \parallel_{4 \over 3} \left( \parallel u
\parallel_{2} \parallel \nabla u \parallel_{2} \right)^{1\over2}
\label{c319}
\\ &\le& \ep  \parallel u \parallel_{2}+  {c_4 \over \ep} \parallel \nabla u
\parallel_{2}   +c_5 \label{c310}
\\
&\le&\ep \left( \parallel u \parallel_{2}+ \parallel \nabla u
\parallel_{2} ^2 \right) +c_6 \label{c321} \\
&\le&\ep \left( \int_{M} { |u|^2 \over{ (1+ |u|)^2}} ~dV_g +
(1+c_2)\parallel \nabla u
\parallel_{2} ^2 \right) +c_7 \label{c322},
\eea
 \noindent
where $ c_3,\cdots c_7$ and $\ep$ are some constants. In the
above, the H\"{o}der inequality, (\ref{c31}) and(\ref{c27}) are
used in (\ref{c318},\ref{c310}), (\ref{c319}) and (\ref{c322}),
respectively.
  There exists a constant $c_8$ such that \bea E(u)&\ge& \int_{M}|\nabla
u|^2 dV_{g} + \int_{M} { |u|^2 \over{ (1+ |u|)^2}} dV_{g} -c_1
\nonumber
\\ &{}& \hbox{\quad} -\ep \left[ \int_{M} {
|u|^2 \over{ (1+ |u|)^2}} ~dV_g + (1+c_2) \int_{M} |\nabla u|^2
~dV_g \right] -c_7
\nonumber \\
&\ge& \left(1- \ep (1+c_2) \right) \int_{M}|\nabla u|^2 dV_{g} +
\left( 1- \ep \right)\int_{M} { |u|^2 \over{ (1+ |u|)^2}} dV_{g}
-c_8 . \label{c331}
 \eea
 \noindent
By taking small $\ep$ and Eq. (\ref{c27}),
 \bea E(u)
 \ge c_9
\Big( \int_{M}|\nabla u|^2 + u^2  dV_{g} \Big)^{1/2} -c_{10} ,\eea
\noindent for some positive constants $c_9$ and $c_{10}$.
\noindent Therefore $E(u)$ is coercive on $H_1^2$ and $\inf_{u\in
H_1^2}E(u)$ is finite. Moreover, $E(u)$  is weakly lower
semi-continuous on $H_1^2$. We take a minimizing sequence
$\{u_n\}$ for $\inf_{u\in H_1^2}E(u)$.  Then $\{u_n\}$ is bounded
on $H_1^2$, which has a subsequence $\{u_{n_k}\}$ converging to
$u\in H_1^2$, a minimizer for $\inf_{u\in H_1^2}E(u)$. By the
elliptic regularity, $u$ is smooth. Finally, $u$ satisfies
Eq.~(\ref{c9}). The existence of a solution for Eq. (\ref{c1}) is
proved.\vskip 0.5 true cm
  Step 2. {\it Behavior of a solution.} \par \noindent
 In this part,  we study the behavior of
the solution of Eq.~(\ref{c1}). Since the distance function is not
smooth, we need to find a smooth function which can tell the
behavior of the solution.

Let $p_0$ be a fixed point in $M$. Take large $R_1>R_0 '$ so that
$\{p_1,\cdots p_n\} \subset B(p_0, R_1)$ and $M-B(p_0, R_1)
\subset M-B(p_0, R_0') ~ \subset C_1 \cup C_2$, where
$C_1=\left(R^2-B_E(0, R_0), h_1 \delta_{ij}\right)$ and
$C_2=\left(R^2-B_E(0, R_0), h_2 \delta_{ij}\right)$  with
$\alpha^{-1}<h_1, h_2<\alpha$.
 Consider $p \in M-B(p_0, R_1)$. Then, either $p \in C_1$ or $C_2$. Using the
 definition of $C_1$ and $C_2$, regard $p$ as a
 point in $\left(R^2-B_E(0,R_0), \delta_{ij}\right) \subset (R^2,
 \delta_{ij}) $. Define $r_e(p)$ be the Euclidean distance from
 $p$ to the origin of $R^2$ and $f_1(p)=-a e^{-b r_e (p)}$
 for $p \in M-B(p_0, R_1) $.
  Note that $f_1 (p)$ is differentiable and
 $\Delta f_1(p)={1 \over {\sqrt {g}}} \Delta_0
 f_1(p)$ for $p \in M-B(p_0, R_1)$ in the canonical coordinates
 system. For any given small $\ep>0$, there exists a constant
 $\delta_1$ so that $ e^t \ge 1-\ep >0$ and $1-e^t\ge (\ep-1) t>0$ for  $
-\delta_1 \le t \le 0$. Take positive constant $b$ so that $b
<(1-\ep)/\sqrt{\alpha}$.

To estimate the lower bound of the solution, we use the Maximum
principle (cf. \cite{Gil}).
 Since $\Delta w \le 0$ and $w\le 0$, we have $ -c
||w||_{L^2(B(p,1))} \le w(p) <0 $
  at  infinity for some positive constant $c$ \cite{Gil}.
 From $\int_M w^2 dV_g < \infty$, $w$ decays to zero uniformly at
infinity.  Since $\alpha^{-1} \delta_{ij} <g_{ij}<\alpha
\delta_{ij} $ on $M- B(p_0, R_1)\subset C_1 \cup C_2$, we can
take sufficiently large $R_2>R_1$ so that there exist positive
constants $a$ and  $\delta_2$ ($<<\delta_1$) with $-\delta_1 \le
f_1(p)=-a e^{-b r_e (p)} \le -\delta_2$ for $p \in \partial
B(p_0,R_2)$ and $w(p)
>- \delta_2 $ for $p \in M- B(p_0,R_2)$.

 Using $\alpha^{-1} \delta_{ij} <g_{ij}<\alpha
\delta_{ij} $ and
 $-\delta_1 <f_1 (p)<0$ for $p \in M-B(p_0, R_2)$,
  \bea \Delta f_1-e^{f_1}
(e^{f_1}-1)&\ge& {1 \over {\sqrt {g}}} \Delta_0 f_1 -(1-\ep)^2 f_1\nonumber\\
&\ge&  {1 \over {\sqrt {g}}}(-{b\over r_e}+b^2)f_1-(1-\ep)^2 f_1\nonumber\\
&\ge&\left( \alpha (b^2-{b\over r_e})-(1-\ep)^2 \right) f_1 \nonumber\\
&>& 0  \label{c-30}.\eea

Take $F(x,t)=e^t(e^t-1)$, then $\partial_t F(x, t) >0$ for
$t>-\delta_1> -\ln 2$. Therefore $F(x, f_1)-F(x, w)=\lambda(x)
(f_1-w)$ for $\lambda(x)>0$ and  $-\delta_1<f_1, w \le 0$. From
(\ref{c-30}), we have the following estimate on $ M-B(p,R_2)$,
 \bea \Delta(f_1 -w) (p)&\ge&
e^{f_1} (e^{f_1} -1)-e^{w} (e^{w} -1)\nonumber \\
&\ge& \lambda(p)(f_1-w). \label{m1}\eea If $f_1-w >0$ on some
domain $D \subset M-B(p, R_2)$, then $\Delta(f_1 -w)
>0$ on $D$. By the Maximum Principle, $f_1 -w$ can not have
maximum inside of $D$. Since $f_1=w$ on $\partial D$ and $f_1>w$
on $D$, $D$ must be the empty set. Therefore, $f_1=-a e^{-b r_e
(p)} \le w$ on $M-B(p_0, R_2)$. Since the metric satisfies
$\alpha^{-1} \delta_{ij}< g_{ij}< \alpha \delta_{ij}$ on $C_1$
and $C_2$, there exists constant $\mu\ge 1$ such that $ \mu^{-1}~
d(p, p_0) <r_e(p) <  \mu~ d(p, p_0)$ for $p \in M-B(p_0, R_2)$.
Finally we conclude that $-a e^{-{b\over \mu} d(p, p_0)} \le w(p)$
on $M-B(p_0, R_2)$.

\vskip 0.3 true cm Next consider decay estimates for $|\nabla w|$.
 Taking partial derivative $\partial_{x_i}$ to Eq. (\ref{c1}),
 we have: \bea
\Delta w_i= e^w (2 e^w-1) w_i \label{w1} \eea \noindent at
infinity where $w_i=\partial_{x_i} w$. Since $e^w \to 1$, $ \Delta
w_i= \zeta (x) w_i $ where $\zeta (x)$ converges to one at
infinity. Similar estimates, as was done for $w$, produce
$|w_i(p)|< a' e^{-{b'\over \mu'} d(p, p_0)}$ for some constants
$a', b'$, and $\mu'$. This completes the proof of Theorem 1.

\section{Conclusion}  \noindent
For an asymptotically flat cylindrical spatial manifold $M$, we
proved the existence and decaying property of a topological
multi-vortex solution of the Chern-Simons Higgs theory in $(2+1)$
space $R\times M$, which have  been previously studied on $(2+1)$
space $R\times R^2$.   The related questions for other prescribed
asymptotic solutions in $(2+1)$ space $R \times M$ with $|\phi|
\to v $ or zero at   infinity of $C_1$ or $C_2$ need future study.

\section*{Acknowledgments}
 \noindent
Author thanks to Yoonbai Kim for  his  help for possible physical
application of this work to recently developed  topics and
interest in this work. This work was supported in part by Faculty
Research Fund Sungkyunkwan University 1999.


\begin{thebibliography}{100}
\bibitem{HKP} J. Hong, Y. Kim, and P. Y. Pac, Phys. Rev. Lett. {\bf 64},
2230 (1990).
\bibitem{JW}
 R. Jackiw and E. J. Weinberg, {\it ibid} {\bf 64}, 2234 (1990).
 \bibitem{RS}  L. Randall and R. Sundrum, Phys. Rev. Lett. {\bf 83}, 4690
(1999).
\bibitem{Gre} R. Gregory, Phys. Rev. Lett. {\bf 84}, 2564 (2000).
\bibitem{MT} D. Mateos and P. K. Townsend, Preprint DAMTP-2001-21,
hep-th/0103030.
\bibitem{Wan} R. Wang, Comm. Math. Phys. {\bf 137}, 587 (1991).
\bibitem{SY} J. Spruck and Y. Yang, Ann. Inst. H. P. {\bf 12}, 75
(1995).
\bibitem{SY2} J. Spruck and Y. Yang, Comm. Math. Phys. {\bf 149}, 361 (1992).
\bibitem{CI} D. Chae and O. Imanuvilov, Comm. Math. Phys. {\bf 215},
119 (2000).
\bibitem{Hf} G.'t Hooft, Nucl. Phys.  {\bf  B 153}, 141 (1979).
\bibitem{CY} L. Caffarelli and Y. Yang, Comm. Math. Phys. {\bf 168}, 321
(1995).
\bibitem{Tar} G. Tarantello, J. Math. Phys. {\bf 37}, 3769 (1996).
\bibitem{DJLW1} W. Ding, J. Jost, J. Li, G. Wang, Calc. Var. PDE {\bf 7}, 87 (1998).
\bibitem{NT} M. Nolasco and G. Tarantello, Calc. Var. {\bf 9}, 31 (1999).
\bibitem{DJLW2} W. Ding, J. Jost, J. Li, G. Wang, Comm. Math. Helv. {\bf
74}, 118 (1999).
\bibitem{DJLW3} W. Ding, J. Jost, J. Li, X. Peng, G. Wang,
Comm. Math. Phys. {\bf 217}, 383 (2001).

\bibitem{Sch} J. Schiff, J. Math. Phys. {\bf 32}, 753 (1991).
\bibitem{Choe} K. Choe, {\it Existence of Solution the Self-dual
Chern-Simons Higgs Theory in a Background Metric}, 2000 Aug., GARC
Preprint, SNU.
\bibitem{KK}S. Kim and Y. Kim, Preprint math-ph/0012045.
\bibitem{FG} W. Fuertes and J. Guilarte,
Eur. Phys. J. C. {\bf 9}, 167 (1999).
\bibitem{ssy} L. Sibner, R. Sibner, Y. Yang,
Proc. R. Soc. Lond. A. {\bf 456}, 593 (2000).
\bibitem{Cle2} G. Clement, {\it Phys. Rev. D} {\bf 54} 1844
(1996).

\bibitem{cckk} B. Chung, J. Chung, S. Kim and Y. Kim,  Preprint
gr-qc/0102104, to appear in Annals of Phys.

\bibitem{Au} T. Aubin, {\it Nonlinear Analysis on Manifolds,
Monge-Ampere Equation} (Springer, New York, 1982).
\bibitem{Yi} Y. Yang,
Comm. Math. Phys. {\bf 186}, 199 (1997).

\bibitem{Gil} D. Gilbarg and N. Trudinger, {\it Elliptic Partial Differential
Equations of Second Order} (Springer, New York, 1983).
\bibitem{jt} A. Jaffe and C. Taubes, {\it Vortices and  Monoploles}
(Birkhauser, Boston, 1980).

\end{thebibliography}
\end{document}